\documentclass[aps,prl,twocolumn,superscriptaddress,showpacs,10pt]{revtex4-2}
\usepackage{amsmath}
\usepackage[dvips,xdvi]{graphicx}
\usepackage{amssymb}
\usepackage{epsfig}
\usepackage{xcolor}
\usepackage{physics}
\usepackage[colorlinks=true,citecolor=blue,urlcolor=magenta,linkcolor=blue]{hyperref}

\begin{document}
	\title{Observation of ergodicity breaking and quantum many-body scars in spinor gases}
	\author{J. O. Austin-Harris}
	\author{I. Rana}
	\author{S. E. Begg}
	\author{C. Binegar}
	\author{T. Bilitewski}
	\email{thomas.bilitewski@okstate.edu}
	\author{Y. Liu}
	\email{yingmei.liu@okstate.edu}
	\affiliation{Department of Physics, Oklahoma State University, Stillwater, Oklahoma 74078, USA}
	\date{\today}
	\begin{abstract}
		We experimentally and theoretically demonstrate spinor gases driven by spin-flopping fields are excellent platforms for investigating ergodicity breaking and quantum scarring. We observe that specific initial states remain nonthermal at weak driving despite the majority of states thermalizing, which constitutes clear evidence of quantum many-body scars (QMBS). As the driving strength increases, the experimental system undergoes a smooth transition from integrable to weakly ergodicity breaking, which supports QMBS, and then to fully thermal.
		This is in agreement with the theoretical spectra, which predict towers of states dissolving with increasing driving strength.   
		This work advances the study of QMBS and quantum scars with applications to, e.g., quantum information storage.
	\end{abstract}
	\maketitle
	
	Understanding how isolated quantum systems thermalize is a topic of immense fundamental interest \cite{Deutsch1991,  Srednicki1994, Rigol2008, Goldstein2010, Polkovnikov2011, Alessio2016, Deutsch2018, Ueda2020} with a wide range of potential applications, e.g.,  quantum transport, quantum metrology, and quantum information storage~\cite{Chien2015, Bluvstein2022, Dooley2023}. The eigenstate thermalization hypothesis (ETH)~\cite{Deutsch1991,Srednicki1994,Deutsch2018}, which predicts most quantum states thermalize to a statistical ensemble, has been successfully used to describe the thermalization dynamics of many quantum systems; however, there are a variety of cases where the ETH fails. 
	Both integrable and many-body localized systems evade thermalization due to the existence of an extensive number of conserved quantities~\cite{Kinoshita2006, Rigol2007, Rigol2009, Serbyn2013, Nandkishore2015, Langen2016, Abanin2019,Jared1,Eisert2015}; whereas in Hilbert space fragmentation~\cite{Khemani2020, Sala2020, Moudgalya2022}, systems avoid thermalization because of constraints that fragment the dynamically accessible Hilbert space into exponentially many disconnected parts. 
	Additionally, systems with quantum many-body scars (QMBS) or quantum scars violate ETH~\cite{Heller1984, Turner2018, Pilatowsky2021, Serbyn2021, Hummel2023, Su2023, Zhang2023, Dag2024, Sinha2024, DagPreprint}.  
	
	QMBS are defined as a thermodynamically small fraction of nonthermal states embedded in a sea of thermal chaotic states of a many-body system, resulting in atypical behaviour for a subset of initial states and weak ergodicity breaking~\cite{Turner2018, Serbyn2021}, which has recently been observed in several experiments \cite{Bernien2017, Turner2018, Bluvstein2022, Su2023, Zhang2023}. 
	Distinctly, quantum scars occur in chaotic systems with a classical phase space when states are scarred by an underlying unstable periodic orbit (UPO), restricting the dynamics to some fraction of the phase space near the UPO; in contrast, unscarred states explore the full phase space~\cite{Heller1984, Pilatowsky2021, Hummel2023, Dag2024, Sinha2024, DagPreprint}.  Despite the similarity in name, a connection between quantum scars and QMBS has yet to be established, in part due to the lack of a classical phase space in many of the systems in which QMBS have been observed~\cite{Pilatowsky2021, Sinha2024, Dag2024, DagPreprint}.
	
	In this paper we experimentally and theoretically study the interplay of quantum scarring, QMBS, and thermalization in spinor Bose-Einstein condensates (BECs). 
	Spinor BECs are highly controllable many-body quantum systems that possess a spin degree of freedom with all-to-all spin interactions and a well-defined phase space~\cite{Jared1,Chang2005,Dag2024,Jared3,Zach1,Lichao2014, Jie2014,Yingmei2009,Lichao2015, Black2007, Stamper2013}. We explore quantum scarring by breaking the conservation of angular momentum with a pair of well-chosen near resonant rotating fields that drive transitions between multiple spin states. The driving fields perturb the classical orbits, causing them to become unstable even for very weak driving strengths $p$, and the resulting UPO scars the majority of energy states~\cite{Dag2024}. Our data demonstrate that these chaotic, scarred states thermalize at all nonzero $p$. In contrast, at sufficiently low $p$, regular states, which lie in the nonchaotic region of phase-space with stable orbits, are QMBS, and consequently we observe long-lived nonthermal plateaus for certain initial states.
	The system is found to smoothly transit from an integrable regime to a weakly ergodicity breaking regime, where QMBS exist, and then to a fully thermal regime as $p$ increases.
	Additionally, states lying on the UPO retain memory of $\eta$, the relative phase between components of nonzero spin.  Our data show that $\eta$ determines both the transient spin dynamics and the long-time equilibrated values. 
	By revealing conclusive signatures of regular and scarred states in many-body spinor gases, this Letter establishes spinor condensates as an ideal testbed to investigate thermalization dynamics of a quantum system with a scarred phase space in the presence of QMBS.
	
	\begin{figure*}[tb]
		\includegraphics[width=176mm]{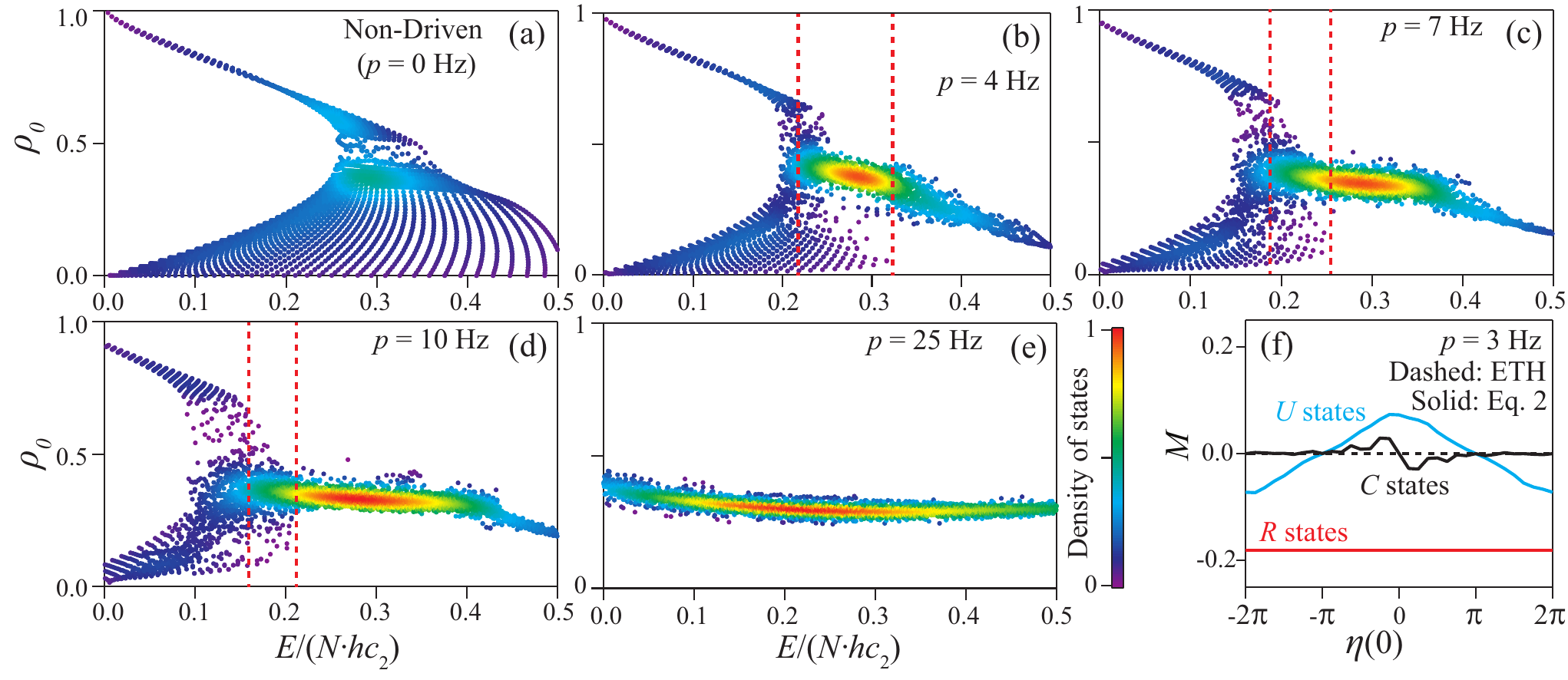}
		\caption{(a-e) Eigenstate expectation values of $\rho_0$ obtained using the first-order Floquet expansion of Eq.~\eqref{Hamiltonian} (see SM~\cite{SM}) versus energy for various driving strengths $p$ crossing the integrable to weakly ergodicity breaking to fully thermal transition. The color scale shows the normalized density of states. Eye-guiding dashed lines indicate an energy-region with both presence of athermal towers of states and thermal states. The towers of states in these regions are QMBS states (see text).  For all initial states studied in this work, $E/(N\cdot hc_2)\approx0.24$ in the $q_B$-infinity limit. (f) Predicted phase dependence of long-time-average ($t\sim 100/c_2$) of $M$ from exact dynamics based on Eq.~\eqref{Hamiltonian}, compared with corresponding ETH predictions (dashed lines). All results for 100 particles with $c_2 = 26$~Hz and $q_B = 41$~Hz.} \label{FigTheory}
	\end{figure*}   
	
	Each experimental cycle begins with an $F=1$ sodium BEC in optical traps under a magnetic field with linear (quadratic) Zeeman energy $h\cdot p_B$ ($h\cdot q_B$), where $h$ is the Planck constant. Microwave and radio frequency (RF) pulses are used to prepare initial states of desired $\rho_0(0)$, $M(0)$, and $\theta(0)$ at time $t=0$. Here $\rho_{m_F}(t)$ is the fractional population in the hyperfine $m_F$ state, $M(t)=\rho_1(t)-\rho_{-1}(t)$ is the magnetization, and $\theta(t)=\theta_{m_F=1}(t) +\theta_{m_F=-1}(t)-2 \theta_{m_F=0}(t)$ is the relative phase among all spin states.  In this work we study three initial states: the \textit{R} state with $\rho_0(0)=0$, $M(0)=-0.7$, and $\theta(0)=0$, chosen to overlap with one tower of regular states; the \textit{U} state with $\rho_0(0)=0.6$, $M(0)=0$, and $\theta(0)=\pi$, chosen to lie on a UPO in phase space; and the \textit{C} state with $\rho_0(0)=0.26$, $M(0)=-0.3$, and $\theta(0)=0$, a chaotic state chosen to have similar per particle energy as the other states in the $q_B$-infinity limit. The \textit{R} and \textit{C} states are Fock states and the \textit{U} state is a spin-coherent state (see SM~\cite{SM}). We then hold the atoms for a time $t_1$ to imprint a desired initial phase $\eta(0)$, where $\eta(t)=\theta_{m_F=1}(t)-\theta_{m_F=-1}(t)$. RF fields of frequencies $\omega _{\pm}=2\pi(p_B\pm q_B)$, are applied at $t=t_1$ for a time $t_2$ to drive $m_F=0\leftrightarrow m_F=\pm1$ transitions and initiate chaotic spin dynamics in the majority of energy states.  The driving fields induce an interaction $\hat{H}_{\mathrm{D}}=h[(\sin((t-t_1)\omega_+) + \sin((t-t_1)\omega_{-}))]\vb{p}\cdot\vb{\hat{S}}$, where $h \cdot |\vb{p}|=h \cdot p$ is the associated linear Zeeman energy and $\vb{\hat{S}}$ is the spin operator. The dynamics are then monitored via spin-resolved imaging~\cite{SM}.
	\begin{figure*}[tb!]
		\includegraphics[width=176mm]{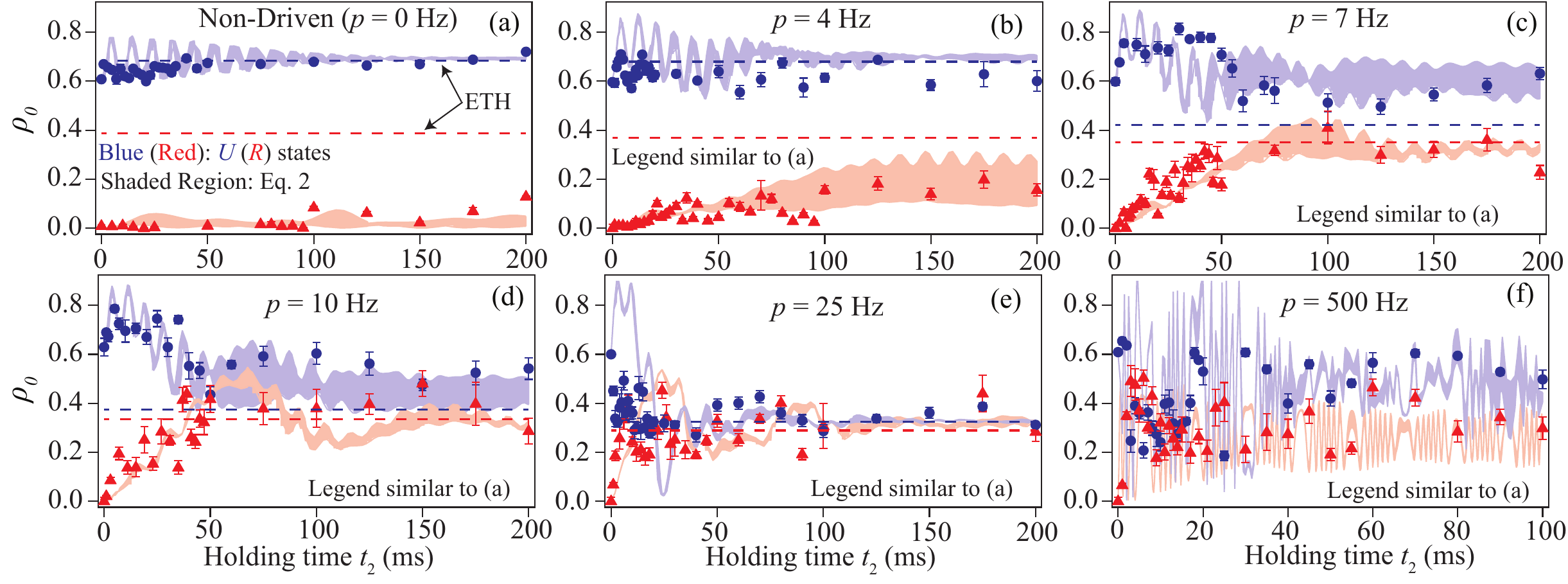}
		\caption{Circles (triangles) display experimental $\rho_0$ time traces taken with the \textit{U} (\textit{R}) initial state when $p=$ (a) $0$, (b) $4$, (c) $7$, (d) $10$, (e) $25$, and (f) $500~\mathrm{Hz}$.  Shaded regions are  Eq.~\eqref{Hamiltonian} predictions with shading indicating $10\%$ uncertainty in $c_2$.  Dashed lines show ETH predictions.} \label{timetrace}
	\end{figure*}
	
	In this study, we observe no spin domains or multiple spatial modes, motivating the description of our observations with Eq.~\eqref{Hamiltonian}, a Hamiltonian derived from a single-spatial mode approximation (SMA) (see SM~\cite{SM}),
	\begin{equation} \label{Hamiltonian}
		\begin{aligned}
			\hat{H} &= \frac{h c_2}{N} \left[ \hat{N}_0 (N - \hat{N}_0 ) + 1/2 (\hat{N}_+ -\hat{N}_-)^2\right]  - h p_{\perp}/2 \, \hat{S}_y  \\
			& + \frac{h c_2}{N} \left(e^{4\pi i  q_B t} \hat{a}_0 \hat{a}_0 \hat{a}_{-}^{\dagger}  \hat{a}_+^{\dagger} + \mathrm{h.c.} \right) \\
			&+h p_{\perp}/(2\sqrt{2}) \left(  i e^{-4\pi i q_B t} \hat{a}_0 \hat{a}_{+}^{\dagger} + i e^{4\pi i q_B t} \hat{a}_0^{\dagger} \hat{a}_{-}  + \mathrm{h.c.}\right)
		\end{aligned}
	\end{equation}
	Here $c_2\sim30~\mathrm{Hz}$ is the spin-dependent interaction, $q_B\approx40~\mathrm{Hz}$, $N$ is the total atom number, the $y$ direction is orthogonal to the quantization axis set by the external magnetic field, $p_{\perp}$ is the magnitude of $\vb{p}$ in the $y$ direction, and $\hat{a}_{m_F}$ $(\hat{a}_{m_F}^\dagger)$ $[\hat{N}_{m_F}]$ is the annihilation (creation) [number] operator for the $m_F$ state~\cite{SM}. 
	
	Spin dynamics under the SMA have been well studied~\cite{Chang2005,Jared3,Zach1,Lichao2014,Jie2014,Yingmei2009,Lichao2015,Black2007, Stamper2013, Evrard2021, Yang2019,Chang2005, Sadler2006, Prufer2018, Huh2024}, in particular for the nondriven case where $p=0$~Hz and the model is fully integrable, both at the classical and quantum level \cite{Lamacraft2011,Bogoliubov2006}.  A nonzero $p$, however, breaks magnetization conservation and introduces chaotic behavior and complex thermalization dynamics in which the majority of energy states are scarred by an underlying UPO in phase space \cite{Dag2024}. In the $q_B$-infinity static limit of Eq.~\eqref{Hamiltonian} it has recently been shown that these chaotic states coexist with towers of regular states that display nonthermalizing behavior and properties typically associated with QMBS \cite{Dag2024}.
	
	To illustrate these features, in Fig.~\ref{FigTheory} we show the eigenstate expectation values of $\rho_0$ versus the energy $E$, for various $p$. As Eq.~\eqref{Hamiltonian} is a time-periodic Hamiltonian with a large frequency $4 \pi q_B$ we consider the first order Floquet high-frequency expansion~\cite{Bukov2015,Eckardt2015,Kuwahara2016} to obtain a static effective Hamiltonian describing the dynamics and thermalization behavior of the system.
	The displayed results are obtained via exact diagonalization of this Floquet Hamiltonian,  valid in the limit $2q_B > c_2 \sim p$.  In the Supplemental Materials~\cite{SM}, we show that this correction crucially needs to be included, and is sufficient to properly describe dynamics and thermalization behaviour for the experimental parameter regime of interest.
	For $p=0$, Fig.~\ref{FigTheory}(a), the system is integrable~\cite{Lamacraft2011,Bogoliubov2006} and the results are highly structured. For $p \geq 25$~Hz, Fig.~\ref{FigTheory}(e),  the distribution of expectation values is narrow and the system obeys strong ETH across the spectrum. For intermediate $p$, Figs.~\ref{FigTheory}(b)-(d), a  tower of regular states  at low densities exists at the same energy as the thermal states~\cite{Dag2024}, which we indicate via dashed lines. These regular states are QMBS states, ensuring the absence of thermalization for any initial state with which they have significant overlap. As $p$ increases, this coexistence region shrinks and the towers dissolve. Notably, our theoretical calculations predict that the QMBS survive in the limit of finite, large $q_B$, i.e., in the first order Floquet-corrected effective Hamiltonian, where the spin-changing collisions remain active. 
	
	Furthermore, Eq.~\eqref{Hamiltonian} predicts that when driving fields with nonzero $p$ are applied, the relative phase $\eta$ between the $m_F=\pm1$ spin components affects both the short-time dynamics, as well as long-time equilibrated observables. 
	Figure~\ref{FigTheory}(f) displays long-time averages for the magnetization $M$ versus the initial phase $\eta(0)$, derived from exact time evolutions under Eq.~\eqref{Hamiltonian} for the three initial states.  The \textit{U} state shows long-time memory of $\eta(0)$ with a strong dependence and athermal values for $M$, and a slightly weaker dependence for $\rho_0$ (not shown). In contrast, the dynamics of \textit{R} states are largely independent of $\eta(0)$, similar to the dynamics of the nondriven system, and retain a large absolute value of $M$ and low $\rho_0$ (not shown) demonstrating absence of thermalization.   Additionally, the \textit{C} state thermalizes to zero magnetization as expected.  
	
	\begin{figure}[b!]
		\includegraphics[width=86mm]{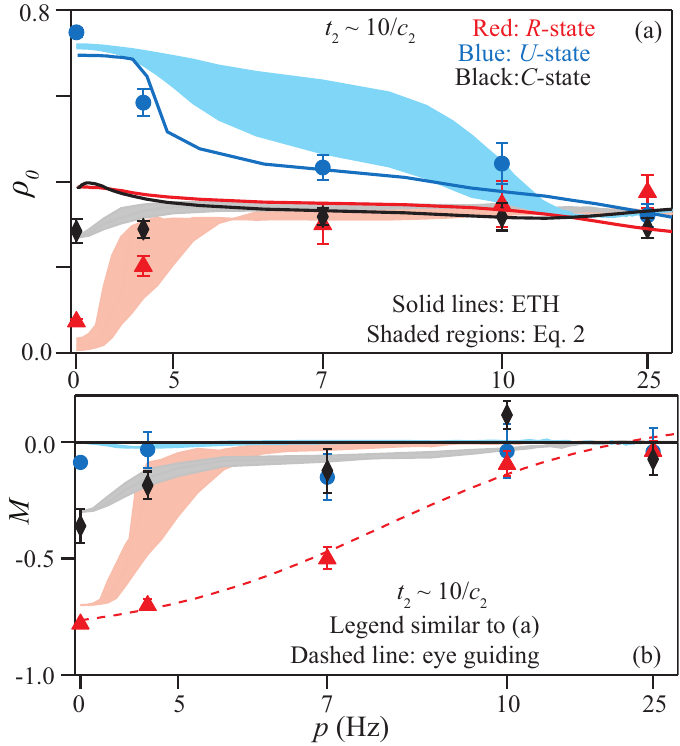}
		\caption{Triangles (circles) [diamonds] display (a) $\rho_0$ and (b) \textit{M} taken with the \textit{R} (\textit{U}) [\textit{C}] initial state at $t=300~\mathrm{ms}\sim10/c_2$, i.e., after the short time dynamics settle out. Shaded regions display Eq.~\eqref{Hamiltonian} predictions  for each initial state accounting for $10\%$ uncertainty in $c_2$, while solid (dashed) lines are ETH predictions (eye-guiding).
		} \label{longhold}
	\end{figure}
	
	The aforementioned theory discussion predicts a clear transition between nonthermal behaviour due to the presence of QMBS at small $p$ to fully thermal behaviour at large $p$, which is directly observable in the dynamics of the chosen initial states.
	A typical example, shown in Fig.~\ref{timetrace}, compares the observed spin dynamics starting from the \textit{R} and \textit{U} states for a range of $p$. The nondriven spinor system, where $p=0$ and $M$ is conserved, is integrable~\cite{Lamacraft2011,Bogoliubov2006} and violates ETH~\cite{Yingmei2009,Jie2014}.  This is confirmed by our observations in Fig.~\ref{timetrace}(a): the observed $\rho_0$ for the \textit{R} state is well-separated from the ETH predictions even after relatively long holding times, as expected for a nonthermalizing state. 
	When $p$ is sufficiently large, however, the system rapidly thermalizes resulting in $\rho_0$ reaching the ETH prediction as shown by the data taken at $p\gtrsim10~\mathrm{Hz}$ (Figs.~\ref{timetrace}(d) and \ref{timetrace}(e)). 
	In between these two regimes, the observed spin dynamics (Figs.~\ref{timetrace}(b) and \ref{timetrace}(c)) smoothly transit between the two extremes with the observed $\rho_0$ remaining clearly distinct from the ETH predictions for the \textit{R} state.  
	Our data also confirms the \textit{R} states remain integrable because spin transitions to the $m_F=1$ state are suppressed leading to an approximate conservation of $M$ (Fig.~\ref{longhold} (b)), which is directly related to the appearance of the tower of states \cite{Dag2024}. As we keep increasing $p$ to the large $p$ limit ($p\gg c_2$), the observed dynamics eventually become dominated by Rabi flopping between the spin states as expected (Fig.~\ref{timetrace}(f)). 
	The observed dynamics can be quantitatively described by predictions derived from Eq.~\eqref{Hamiltonian}, which are displayed as shaded regions in Fig.~\ref{timetrace} to reflect an estimated 10\% uncertainty in the spin-dependent interaction $c_2$. 
	
	The aforementioned smooth transition, from an integrable regime to a weakly ergodicity breaking regime where QMBS exist and then to a fully thermal regime as $p$ increases, is best visualized in long-time expectation values. In Fig.~\ref{longhold} we plot the observed dynamics at a long holding $t_2=300~\mathrm{ms}\sim 10/c_2$. The selection of this  $t_2$ value is suggested by both the time traces shown in Fig.~\ref{timetrace} and our theoretical calculations which illustrate the transient spin dynamics have largely settled out by this point (see SM~\cite{SM}). Figure~\ref{longhold} clearly demonstrates that the \textit{R} state undergoes a transition from nonthermalizing to thermalizing as $p$ is increased. The experimental observations are well described by Eq.~\eqref{Hamiltonian} when the uncertainty in $c_2$ is taken into account (see shaded regions in Fig.~\ref{longhold}). 
	The theory-experiment agreement indicates the \textit{R}  states inside the transition region ($0\lesssim p \lesssim c_2/3$) are QMBS as predicted by Fig.~\ref{FigTheory}. A unique feature of the spinor many-body system, in contrast to other systems that display QMBS~\cite{Turner2018,Serbyn2021,Hummel2023,Pilatowsky2021,Dag2024,DagPreprint,Su2023,Zhang2023}, is that it hosts not only QMBS, but also quantum scars in the form of an underlying UPO in phase space that scars the vast majority of eigenstates. When $p$ is nonzero, one such scarred state is the \textit{C} state, a chaotic state with roughly the same average energy per particle as the \textit{R} and \textit{U} states. The observed spin dynamics (Fig.~\ref{longhold}) confirm that the \textit{C} state thermalizes for even the smallest nonzero $p$ studied. 
	
	We extend the above experiment to a much longer timescale, $t_2=4~\mathrm{s}\sim100/c_2$, as theoretical calculations for the ideal model predict persistence of the nonthermalizing dynamics for considerably long times (see SM~\cite{SM}).  However, our observations, shown in SM~\cite{SM}, indicate that over these timescales the experimental system experiences further relaxation due to energy dissipation~\cite{Jie2014} and additional relaxation channels not modelled here. 
	
	\begin{figure}[tb!]
		\includegraphics[width=86mm]{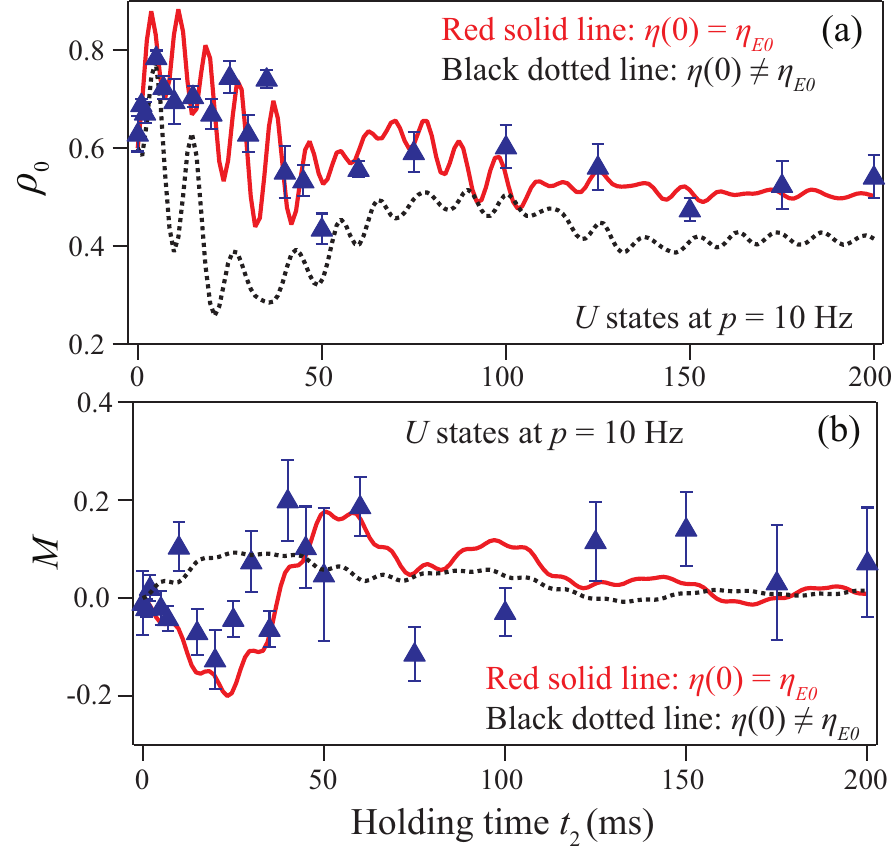}
		\caption{Triangles display the observed (a) $\rho_0$ and (b) $M$ time traces starting from the \textit{U} state when $p=10~\mathrm{Hz}$. Red (black) lines are Eq.~\eqref{Hamiltonian} predictions for a phase $\eta(0)=\eta_{E0}~(\neq\eta_{E0})$ where $\eta_{E0}\sim5\pi/4$ is experimentally imprinted (see SM).} \label{phase}
	\end{figure}
	
	Other major differences are observed between spin dynamics of the driven and nondriven systems. Firstly, our observations show $\eta$ is a new tuning parameter to determine the transient spin dynamics and the long-time equilibrated values of driven \textit{U} states, whereas the nondriven system is independent of $\eta$. The \textit{U} state dynamics are thus capable of retaining memory of the phase $\eta$ (see Fig.~\ref{FigTheory}(f)).
	A typical example highlighting the importance of $\eta$ is shown in Fig.~\ref{phase}: the observed \textit{U} state dynamics are only captured by predictions derived from Eq.~\eqref{Hamiltonian} for $\eta(0)=\eta_{E0}\sim5\pi/4$ as imprinted by our experimental sequence, while predictions for an arbitrary $\eta(0)\neq\eta_{E0}$ significantly deviates from the observations. Good theory-experimental agreement is found for all $p$ values in this work using the experimental $\eta_{E0}\sim5\pi/4$ (see Fig.~\ref{timetrace}, Fig.~\ref{longhold}, and SM~\cite{SM}).
	Secondly, the driving appears to induce large variances in $\rho_0$ and $M$ after certain holding times (Fig.~\ref{timetrace} and Fig.~\ref{phase}). Several effects could contribute to this increasing variance and its underlying physics requires further study, but it may be related to quantum scarring which was recently predicted to be present even for chaotic states and requires temporal averages for ergodicity~\cite{Pilatowsky2021}.
	
	Our work establishes spinor condensates driven by weak spin-flopping fields as excellent platforms for the study of quantum scarring.  In our system, towers of regular states remain nonthermal at weak driving despite the bulk of the states thermalizing to the ETH prediction, demonstrating conclusive signatures of QMBS. By tuning $p$, we observe a smooth transition of the system from integrable to weakly ergodicity breaking, which supports QMBS, and then to fully thermal. Our system additionally hosts an underlying UPO scarring the phase space. While our results reveal that states on the UPO retain memory of the initial phase $\eta$, definitive experimental signatures of  quantum scars associated with the UPO require further study. One potential method is to significantly reduce the atom number by applying optical lattices~\cite{Zihe2019, Jared2021}, which enables the detection of revivals in spin dynamics that are predicted as a signature of quantum scars (see SM~\cite{SM}).  This work opens a new avenue for the study of QMBS, quantum scars, and regular states with potential applications to, for example, quantum information storage.

	\begin{acknowledgments}
		\noindent{\it Acknowledgments --} We thank C. B. Dag for bringing this topic to our attention. We acknowledge support from the Noble Foundation and National Science Foundation through Grant No. PHY-2207777. Some of the computing for this project was performed at the High Performance Computing Center at Oklahoma State University supported in part through the National Science Foundation grant OAC-1531128.
	\end{acknowledgments}

\end{document}


\title{Supplemental Material for ``Observation of ergodicity breaking and quantum many-body scars in spinor gases''}
	\author{J. O. Austin-Harris}
	\author{I. Rana}
	\author{S. E. Begg}
	\author{C. Binegar}
	\author{T. Bilitewski}
	\email{thomas.bilitewski@okstate.edu}
	\author{Y. Liu}
	\email{yingmei.liu@okstate.edu}
	\affiliation{Department of Physics, Oklahoma State University, Stillwater, Oklahoma 74078, USA}
	\maketitle
	\vspace{-2pc}
	\section{Experimental Details}
	Each experimental cycle begins with creating an $F=1$ sodium ($^{23}$Na) BEC of up to $10^5$ atoms in a crossed optical dipole trap with trapping frequencies $\omega_{x,y,z}\approx2\pi\times(140, 140, 180)~\mathrm{Hz}$ under a fixed magnetic field of the linear (quadratic) Zeeman energy $h\cdot p_B$ ($h\cdot q_B$), where $h$ is the Planck constant.  For the experiments presented in this work $p_B\approx270~\mathrm{kHz}$ and $q_B\approx41~\mathrm{Hz}$.  An initial state of desired $\rho_0(0)$, $M(0)$, and $\theta(0)$ is then prepared using resonant microwave and radio frequency (RF) pulses.        
	Here $\rho_{m_F}(t)$ is the fractional population in the hyperfine $m_F$ state, $M(t)=\rho_1(t)-\rho_{-1}(t)$ is the magnetization, and $\theta(t) = \theta_{m_F=1}(t) +\theta_{m_F=-1}(t)-2 \theta_{m_F=0}(t)$ is the relative phase among all spin states.  In this work we study three initial states: the \textit{R} state with $\rho_0=0$, $M(0)=-0.7$, and $\theta(0)=0$, which is chosen to overlap with one tower of regular states; the \textit{U} state with $\rho_0=0.6$, $M(0)=0$, and $\theta(0)=\pi$, which is chosen to lie on an unstable periodic orbit in phase space; or the \textit{C} state with $\rho_0=0.26$, $M(0)=-0.3$, and $\theta(0)=0$, a chaotic state chosen to have similar per particle energy as the other states in the $q_B$-infinity limit. %
	The \textit{C} and \textit{R} state are Fock states defined as $\prod_{m_F}\frac{1}{\sqrt{N_{m_F}!}}  (\hat{a}^{\dagger}_{m_F})^{N_{m_F}} \lvert 0 \rangle$. The \textit{U} state is a spin-coherent state defined as $\frac{1}{\sqrt{N!}} (\sum_{m_F} \sqrt{\rho_{m_F}} e^{i\theta_{m_F}} \hat{a}^{\dagger}_{m_F})^N \lvert 0 \rangle$, where $N$ is the total number of bosons, $N_{m_F} = N \rho_{m_F}$, $\hat{a}^{\dagger}_{m_F}$ is the creation operator for a boson in the $m_F$ hyperfine level and $\lvert 0 \rangle$ is the vacuum state. %
	Due to technical differences in the state preparation, the atom number after preparing a \textit{U} state is consistently much higher than the atom number after preparing \textit{R} or \textit{C} states. Correspondingly, the average spin dependent interaction $c_2$ is higher for \textit{U} states $(c_2\approx35~\mathrm{Hz})$, than it is for \textit{R} and \textit{C} states $(c_2\approx25~\mathrm{Hz})$. The nominal value listed in the main text $(c_2\sim30~\mathrm{Hz})$ represents an average of these two conditions and the difference has been accounted for in all theoretical calculations in this work. After the initial state preparation, we hold the atoms for a short time $t_1$ to imprint a desired $\eta$, where $\eta(t)=\theta_{m_F=1}(t)-\theta_{m_F=-1}(t)$ is the relative phase between components of nonzero spin. Both $\eta$ and $\theta$ evolve during $t_1$, however at much different timescales: a change in $\eta$ of $2\pi$ requires a change in $t_1$ of approximately $4~\mu\mathrm{s}$ determined by $p_B$, while a change of $2\pi$ in $\theta$ requires a change in $t_1$ of approximately $15~\mathrm{ms}$ determined by $q_B$ and $c_2$~\cite{Jared3,Lichao2014}. In this work,  $t_1$ is a constant, up to an integer multiple of $1/p_B$. The experimental $t_1$ is on the order of microseconds, and $\theta$ can therefore be considered to remain constant during the short holding. A pair of near resonant RF fields of frequencies $\omega _{\pm}$, where $\omega_\pm=2\pi(p_B\pm q_B)$, are then applied at $t=t_1$ for a time $t_2$ to drive $m_F=0\leftrightarrow m_F=\pm1$ spin transitions and initiate chaotic spin dynamics in the majority of energy states, while the atoms are held in the optical trap.  The driving fields induce an interaction of the form $\hat{H}_{\mathrm{D}}=h[(\sin((t-t_1)\omega_+) + \sin((t-t_1)\omega_{-}))]\vb{p}\cdot\vb{\hat{S}}$ where $t>t_1$ and $h\cdot |\vb{p}|= h \cdot p$ is the linear Zeeman energy induced by the driving fields. The driving-induced quadratic Zeeman energy can be neglected due to its much smaller energy scale. The atoms are then abruptly released for time of flight expansion and Stern-Gerlach spin state separation prior to standard absorption imaging.
	
	\begin{figure}[b!]
		\includegraphics[width=160mm]{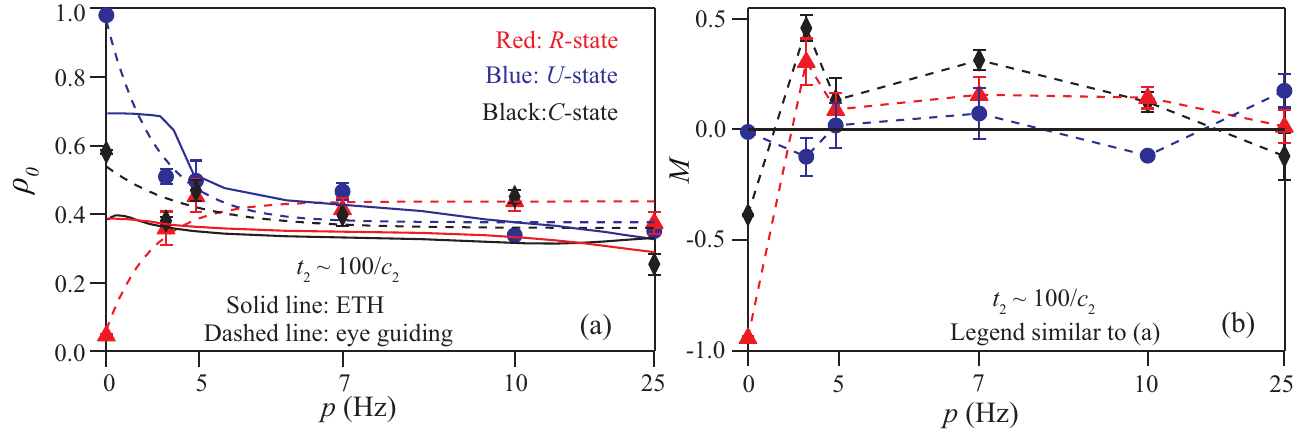}
		\caption{Triangles (circles) [diamonds] display the observed (a) $\rho_0$ and (b) \textit{M} taken with the \textit{R} (\textit{U}) [\textit{C}] initial state at an extremely long holding time, $t_2=4~\mathrm{s}\sim100/c_2$. Solid (dashed) lines are ETH predictions (guide the eye).
		} \label{4s}
	\end{figure}

	In Fig.~\ref{4s}, we extend the experiment shown in Fig.~\ref{longhold} of the main text to a much longer timescale, $t_2=4~\mathrm{s}\sim100/c_2$, as theoretical calculations for the ideal model predict persistence of the non-thermalizing dynamics for considerably long times. Our data in this figure indicate that over these timescales the experimental system experiences further relaxation due to energy dissipation~\cite{Jie2014} and additional relaxation channels not modelled in this work.  This and other effects associated with long holding complicate the physical interpretation of the observed dynamics, although the experimental $\rho_0$ and $M$ of the driven system ($p\neq 0$) show some agreement with ETH predictions (solid lines in Fig.~\ref{4s}).  However, the fundamentally different nature induced by the driving is reinforced by the drastic differences observed in Fig.~\ref{4s} at zero and nonzero $p$: for example, $M$ is somewhat conserved (not conserved) for the nondriven (driven) system.

	\section{Details on theoretical modelling}
	
	\subsection{Derivation of System Hamiltonian}
	In this section we derive Eq.~\eqref{Hamiltonian} of the main text, which follows the discussion in the Supplemental Material of Ref. \cite{Dag2024} with some modifications due to experimental details. In the following we set $h=1$.
	
	We consider the spinor gas under the single spatial mode approximation (SMA)~\cite{Ohmi1998, Law1998, Black2007}. The total Hamiltonian in the lab frame is given by
	\begin{equation}
		\hat{H}_{\mathrm{lab}} = \hat{H}_{Z} + \hat{H}_D + \hat{H}_S.
	\end{equation}
	
	The applied magnetic (Zeeman) field pointing along the $\bm{q}$ direction  (where $|\bm{q}|=1$) is given by
	\begin{align}
		\hat{H}_Z = p_B (\bm{q} \cdot\bm{\hat{s}_i})  + q_B \sum_i(\bm{q} \cdot\bm{\hat{s}_i})^2,
	\end{align}
	where $p_B \approx 270 ~\text{kHz}$ and $q_B \approx 41~\text{Hz}$. The additional driving fields applied along the  $\bm{p}$ axis ($|\bm{p}|=1$) are described by
	\begin{align}
		\hat{H}_D =  p~[\sin (\omega_+ t ) + \sin(\omega_-  t)]\bm{p}\cdot\bm{\hat{S}}
	\end{align}
	where $\omega_{\pm} = 2 \pi( p_B \pm q_B)$ is chosen in order to induce resonant transitions between spin states. In the lab frame $\bm{p} = (0,0,1)$ and $\bm{q} \approx(0.9, 0.1, 0.4)$.

	Finally, the spin-spin interactions are
	\begin{align}
		\hat{H}_S= \frac{c_2}{2N} \bm{\hat{S}}^2 = \frac{c_2}{2N}\big[ 2 \hat{N}_0 (\hat{N}_+ + \hat{N}_-) + \hat{N}_+^2 - 2 \hat{N}_+\hat{N}_- + \hat{N}^2_- + 2 (\hat{a}_0^2 \hat{a}^{\dagger}_+\hat{a}^{\dagger}_- + \hat{a}_0^2 \hat{a}_+ \hat{a}_-) + 2N - \hat{N}_+ - \hat{N}_- \big] ,
	\end{align}

	We first rotate into the frame where $\hat{H}_Z$ is diagonal, which is equivalent to a choice of quantization axis aligned with the direction $\bf{q}$.
	This yields 
	\begin{align}
		H_{Z,\text{diag}} = p_B \hat{S}_z + q_B \sum_i (\hat{s}^z_i)^2,
	\end{align}
	with the single-particle eigenenergies $(-p_B+q_B,0,p_B+q_B)$. The drive is transformed into $\hat{H}_R =  p~[\sin (\omega_+ t ) + \sin(\omega_-  t)](R\bm{p})\cdot\bm{\hat{S}}$ with a rotation matrix $R$ that rotates $\bf{q}$ into $(0,0,1)$. We can choose this such that
	\begin{equation}
		\hat{H}_D =  p_{\parallel}~[\sin (\omega_+ t ) + \sin(\omega_-  t)] \hat{S}_z + p_{\perp}~[\sin (\omega_+ t ) + \sin(\omega_-  t)] \hat{S}_x 
	\end{equation}
	where $p_{\parallel}$ and $p_{\perp}$ are respectively the parallel and orthogonal components of $\bf{p}$ with respect to $\bf{q}$. This makes apparent that $p_{\perp}$ drives transitions between the single-particle eigenstates, whereas $p_{\parallel}$ results in a minor time-dependent modulation of the energies. The spin interaction $\hat{H}_S$ corresponds to the total spin length and is therefore invariant under rotations.

	We now rotate out $\hat{H}_{Z,\text{diag}}$ by moving to the interaction picture. This alters the spin-spin interactions, with time-dependent factors appearing in the spin collision terms:
	\begin{align}
		\hat{H}'_S &= \frac{c_2}{2N}\big[ 2 \hat{N}_0 (\hat{N}_+ + \hat{N}_-) + \hat{N}_+^2 - 2 \hat{N}_+\hat{N}_- + \hat{N}^2_- + 2 (e^{i 4 \pi q_Bt}\hat{a}^{\dagger}_+\hat{a}^{\dagger}_-  \hat{a}_0^2 + e^{-i4\pi q_Bt} \hat{a}^{\dagger 2}_0 \hat{a}_+ \hat{a}_-) + 2N - \hat{N}_+ - \hat{N}_- \big] \nonumber.
	\end{align}
	Similarly, the new drive term is
	\begin{align}
		\hat{H}'_{D}  =  - \frac{p_{\perp}}{2} \, \hat{S}_y  + \frac{p_{\perp}}{2\sqrt{2}} \left(  i e^{-4\pi i q_B t} \hat{a}_0 \hat{a}_{+}^{\dagger} + i e^{4\pi i q_B t} \hat{a}_0^{\dagger} \hat{a}_{-}  + \mathrm{h.c.}\right), 
	\end{align}
	where we dropped terms rotating with $p_B$, but retained $q_B$-dependent terms, consistent with the hierarchy $p_B \gg q_B$. 
	
	The total Hamiltonian $\hat{H} = \hat{H}_S' + \hat{H}_D'$, which appears in  main text Eq.~\eqref{Hamiltonian}, is therefore
	\begin{equation}
		\begin{aligned}
			\hat{H}  &= \frac{c_2}{N} \left[ \hat{N}_0 (N - \hat{N}_0 ) + 1/2 (\hat{N}_+ -\hat{N}_-)^2\right]  - \frac{p_{\perp}}{2} \, \hat{S}_y  \\
			& + \frac{c_2}{N} \left(e^{4\pi i  q_B t} \hat{a}_0 \hat{a}_0 \hat{a}_{-}^{\dagger}  \hat{a}_+^{\dagger} + \mathrm{h.c.} \right) +\frac{p_{\perp}}{2\sqrt{2}} \left(  i e^{-4\pi i q_B t} \hat{a}_0 \hat{a}_{+}^{\dagger} + i e^{4\pi i q_B t} \hat{a}_0^{\dagger} \hat{a}_{-}  + \mathrm{h.c.}\right) .
		\end{aligned} \label{SI-eq:Ht}
	\end{equation}
	Assuming $q_B\gg c_2\gg p$ and taking the limit $q_B \rightarrow \infty$, we obtain the static Hamiltonian 
	\begin{equation}
		\begin{aligned}
			\hat{H}_{q_B \rightarrow\infty}  &= \frac{c_2}{N} \left[ \hat{N}_0 (N - \hat{N}_0 ) + 1/2 (\hat{N}_+ -\hat{N}_-)^2\right]  - \frac{p_{\perp}}{2} \, \hat{S}_y  ,
		\end{aligned} \label{eq:staticHam}
	\end{equation}
	which corresponds to the Hamiltonian obtained in Ref.~\cite{Dag2024}, albeit with a drive in the y-direction rather than the x-direction.

	\subsection{Derivation of Floquet Corrections}
	
	\begin{figure}[tb]
		\includegraphics[width=0.495\textwidth]{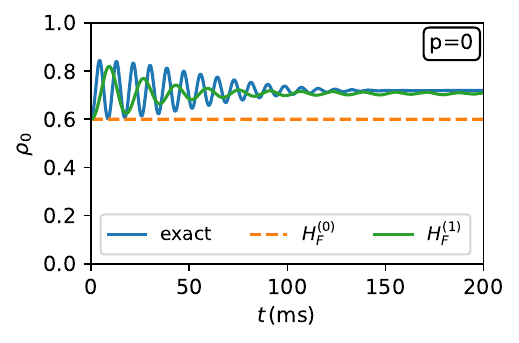}
		\includegraphics[width=0.495\textwidth]{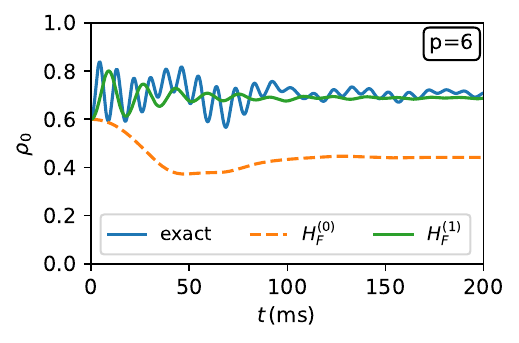}
		\caption{Dynamics of the fractional population $\rho_0$ for the $\textit{U} $ state for $p=0$ Hz (left) and $p=6$ Hz (right) at  $\eta(0)=0$ for $N=100$, $q=40$ Hz and $c_2=40$ Hz. We compare the exact dynamics under the time-dependent Hamiltonian, with the static approximation $\hat{H}_F^{(0)}$ and including the first-order correction $\hat{H}_F^{(1)}$. \label{SI-fig:Floquet_correction_dyn}}
	\end{figure}
	
	While the experiment is not operating in the limit $q_B \rightarrow \infty$, we still have $q_B > c_2$ and $q_B>p$ (at least in the region of interest where the transition happens). Thus, the dynamics under the time-periodic Hamiltonian, Eq.~\eqref{SI-eq:Ht}, with period $T=2 \pi/\Omega$ and frequency $\Omega= 2 q_B/\hbar$ may in this regime be described by an effective time-independent Hamiltonian $\hat{H}_F$ which can be expanded in powers of the inverse frequency \cite{Bukov2015, Eckardt2015,Kuwahara2016}.
	
	Taking the Floquet-Magnus expansion at first order, $\hat{H}_F \approx \hat{H}_F^{(0)} + \hat{H}_F^{(1)} $. $ \hat{H}_F^{(0)}$ is just the average static Hamiltonian, Eq.~\eqref{eq:staticHam}, and the second order term is 
	\begin{equation}
		\hat{H}_F^{(1)} = \frac{1}{2 T i \hbar} \int_{0}^{T} dt_1 \int_{0}^{t_1} dt_2 [H(t_1), H(t_2)]  
	\end{equation}
	which turns out to be
	\begin{align}
		\hat{H}_F^{(1)}&= \frac{1}{2q_B} \Bigg[-\frac{c_2^2}{N^2}  \Big(\hat{N}_0^2\hat{N}_+-4 (\hat{N}_-\hat{N}_0\hat{N}_++\hat{N}_0\hat{O}_{-0}\hat{O}_{+0}+\hat{O}_{0-}\hat{O}_{0+}\hat{N}_0+ \hat{O}_{-0} \hat{O}_{+0}) \nonumber \\
		&\quad -\hat{N}_0\hat{N}_+-2 \hat{N}_-\hat{N}_++\hat{N}_-\hat{N}_0^2 - \hat{N}_-\hat{N}_0+(\hat{N}_0-1) \hat{N}_0+2 N \hat{O}_{-0}\hat{O}_{+0}+2 (N-2) \hat{O}_{0-}\hat{O}_{0+}\Big)  \nonumber  \\
		&+\frac{i c_2 p}{4 \sqrt{2} N} \Big(2   (-\hat{O}_{+0}\hat{N}_++\hat{N}_+\hat{O}_{0+}-\hat{O}_{0-}\hat{N}_++\hat{O}_{-0}\hat{N}_+-\hat{N}_0 \hat{O}_{+0}+\hat{O}_{0+}\hat{N}_0-\hat{O}_{0-}\hat{N}_0+\hat{N}_0\hat{O}_{-0}  \nonumber \\
		& \quad -\hat{N}_-\hat{O}_{+0}+\hat{N}_-\hat{O}_{0+}-\hat{N}_-\hat{O}_{0-}+\hat{O}_{-0}\hat{N}_-)+(2 N-3)   (\hat{O}_{+0}-\hat{O}_{0+}+\hat{O}_{0-}-\hat{O}_{-0}) \Big)  \nonumber  \\
		& +\frac{1}{8} p^2 (\hat{N}_+-2   \hat{N}_0+\hat{N}_--\hat{O}_{+-}-\hat{O}_{-+}) \Bigg] \label{eq:correction}
	\end{align}
	where $\hat{O}_{\alpha, \beta} = \hat{b}^{\dagger}_{\alpha} b_{\beta}$. This provides corrections at order $\frac{p}{2q_B}$ ($\frac{c_2}{ 2q_B}$) to the static Hamiltonian. We  note that we do not expect this approximation to be good beyond $p \sim 2q_B$.
	
	Most notably, it restores coupling between different spin components via the interactions as seen in the terms proportional to $c_2^2/(2q_B)$, which are crucial to capture the $\rho_0$ dynamics correctly, in particular in the weak driving (small $p$) limit. We demonstrate in Fig.~\ref{SI-fig:Floquet_correction_dyn} that this correction captures the dominant difference between the static Hamiltonian and the exact dynamics for the $p$-range of interest in the example of the UPO state. Most importantly, even at $p=0$ the first order correction restores $\rho_0$ dynamics which are absent in the static Hamiltonian in the $q$-infinity limit. The first order correction also significantly improves the long-time equilibrated values during the dynamics compared to the infinite frequency static Hamiltonian. The good quantitative agreement between the exact dynamics and the first-order Floquet Hamiltonian motivates its use to obtain the results for the spectra shown in the main text Fig.~\ref{FigTheory}.
	
	\subsection{Discussion of and comparison with static limit}
	\begin{figure*}[tb]
		\includegraphics[width=176mm]{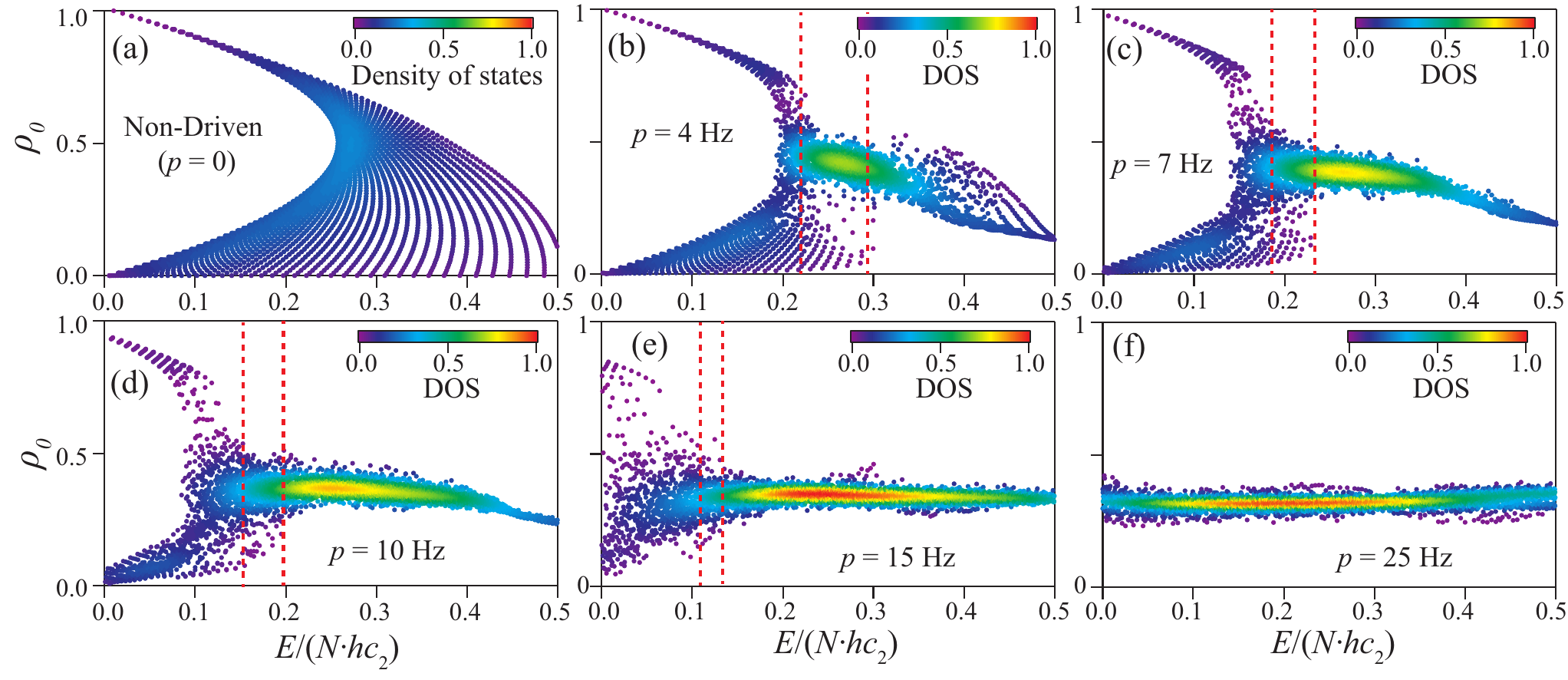}
		\caption{(a-f) Eigenstate expectation values of $\rho_0$ obtained using the static Hamiltonian (Eq.~\eqref{eq:staticHam})  versus energy for various driving strengths $p$ crossing the integrable to weakly ergodicity breaking to fully thermal transition. The color scale shows density of states. Eye-guiding dashed lines indicate an energy-region with both presence of athermal towers of states and thermal states.  The regular, coherent states in these regions are QMBS states. All results for 100 particles with $c_2 = 26~$Hz in the $q_B$-infinity limit. Colorbar shows density of states (DOS). DOS in all panels is normalized to the maximal DOS across all datasets.} \label{FigStaticTheory}
	\end{figure*}
\begin{figure}[tb]
		\includegraphics[width=0.9\textwidth]{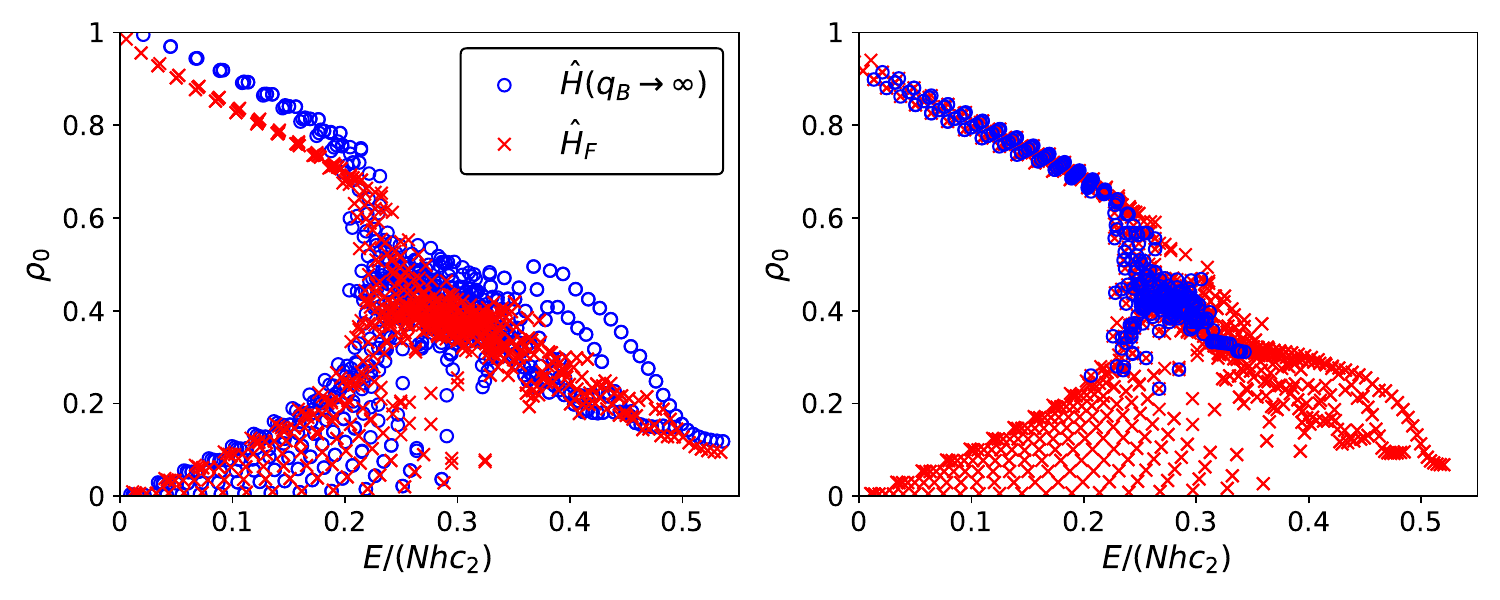}
		\caption{Left: Eigenstate expectation values of $\rho_0$ as a function of energy $E$ for the static Hamiltonian without (crosses) and with (circles) Floquet corrections. Data is for $40$ particles with $p = 3.8~$Hz, $c_2 = 26~$Hz, $
			q_B  = 41~$Hz. Right: similar, but for the Floquet corrected model with 50 particles (crosses) and $p = 3~$Hz, $c_2 = 37.7~$Hz, $q_B  = 41~$Hz. Eigenstates with significant overlap with U states, satisfying $|\braket{v}{\text{U}}|^2 > 10^{-4}$, are circled. }
		\label{fig:stat_vs_floq} 	
	\end{figure}

	In this section we compare eigenstate expectation values for the static Hamiltonian $\hat{H}^{(0)}_F=\hat{H}(q_B \rightarrow \infty$), Eq.~\eqref{eq:staticHam}, against results obtained
	with the Floquet corrected Hamiltonian $\hat{H}_F$ (see Eq.~\eqref{eq:correction} for the correction). This confirms that the phenomenology predicted for the infinite-q limit \cite{Dag2024} persists in the finite-q limit in presence of the Floquet corrections. In Fig. \ref{FigStaticTheory} (a-f) we show $\rho_0$ as a function of energy for different $p$ values, in analogy to the results shown in Fig. \ref{FigTheory} of the main text. The similarity between these two figures indicates that the results are not qualitatively altered by the inclusion of the Floquet correction. To probe this in more detail, Fig. \ref{fig:stat_vs_floq}(left) shows similar results for $p = 3.8$ Hz, directly comparing results obtained with the static Hamiltonian against those when Floquet corrections are included. It can be seen that the models are qualitatively similar, with the correction providing a small shift in the position of the towers of states. 
	
	\subsection{Thermal Values}
	We calculate thermal values for observables $\hat{O}$ at a given energy $E$, determined by the initial state, using the microcanonical average. This is  an average over states in a narrow energy window, $\delta_E = [E - \delta E, E + \delta E]$, given by  
	\begin{align}
		O_{MC}(E) =  \frac{1}{N_E}
		\sum_{\delta_E}
		\bra{v(E)} \hat{O} \ket{v(E)} 
	\end{align}
	where $N_E$ is the number of states in the energy window and $\ket{v(E)}$ are eigenstates at energy $E$. In practise we set  $\delta E$ to the standard deviation of the energy: 
	$\delta E  = \sqrt{\text{Var}[\bra{\psi(0)}\hat{H}\ket{\psi(0)}]}$.
	This works well except at low energies, where for small $p$ values the states fall into two separate groups with low density $\rho_0$ and high $\rho_0$ respectively [see Fig. \ref{FigTheory}(a-d)]. This is due to the energy depending on $\rho_0$ as $E \sim \rho_0(1-\rho_0)$.  In general, the initial states we consider do not lie in this energy range. The exception is the U state at  drive strengths $p \lesssim  20$ Hz. However, due to the relatively high $\rho_0$ value of this state, $\rho_0 \sim 0.6$, it has very little overlap with states in the low density branch and we can expect it to thermalize to the thermal expectation value of the high density branch. To demonstrate this clearly, Fig. \ref{fig:stat_vs_floq}(right) shows eigenstate expectation values of the model with Floquet corrections, with states that have a significant (non-vanishing) overlap with the U state ($\ket{U}$) circled, as determined by the criterion $|\braket{v}{\text{U}}|^2 > 10^{-4}$. It can be seen that there is perhaps only one state from the lower branch that contributes to the U state, while a non-vanishing overlap is observed for a large number of states in the high density branch. To evaluate thermal values for the energy range in question we therefore restrict the microcanonical average to only include states with densities that lie above those of the lower branch states, i.e. $\rho_0 \gtrsim 0.4.$ As shown in Fig. \ref{longhold}, this yields good agreement with the long time time-average of the full model (Eq.~\eqref{Hamiltonian}) and the experimental values, respectively.

	\subsection{Equilibration Dynamics}
\begin{figure}[tb]
		\includegraphics[width=0.95\textwidth]{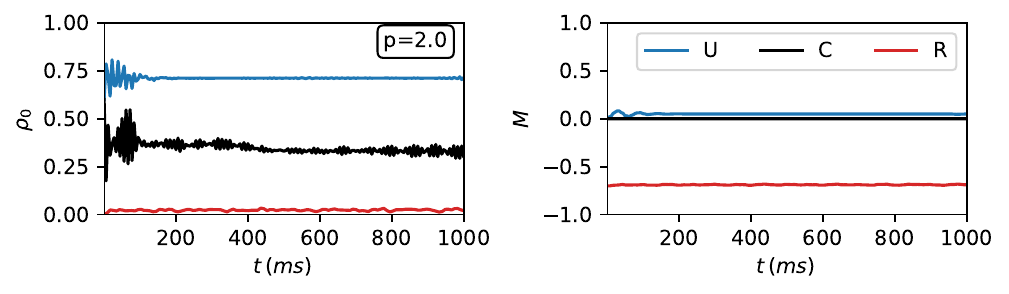}
		\caption{Dynamics of the fractional population $\rho_0$ (left) and the magnetisation $M$ (right) for the  $\textit{R} $,  $\textit{C} $ and  $\textit{U} $ states for $p=2.0$ Hz and $\eta(0)=0$ for $N=100$ . \label{SI-fig:equilibration}}
	\end{figure}

	We extract the theoretical equilibrated values in Fig.~\ref{FigTheory} (f) from long-time exact dynamics of the time-dependent Hamiltonian. We show in Fig.~\ref{SI-fig:equilibration} typical time-traces for the fractional population $\rho_0$ and the magnetisation $M$ for the \textit{R}, \textit{C} and \textit{U} states for small $p=2$ Hz in the athermal regime. We clearly observe that these equilibrate on a time-scale of about 200 ms, and do not show significant late time dynamics, with the exception of the chaotic state that seems to relax closer to about 500 ms. We also note that for $\eta(0)=0$ the $\textit{U}$ state clearly equilibrates to a positive magnetisation (with the $\eta$ dependence shown in Fig.~\ref{FigTheory} (f) of the main text), whereas the chaotic state equilibrates to zero magnetisation, and the regular state retains its large negative initial magnetisation showing absence of thermalisation. More specifically, the initial phase $\eta(0)$ determines whether the magnetisation for the $\textit{U}$ state is initially increasing or decreasing. Finally, we note that the equilibration time does depend both on the specific initial state, as well as the parameters, and we choose times that ensure there is a clear plateau to extract the long-time equilibrated values.
	
	\subsection{Revivals for the UPO state}
	\begin{figure}[tb]
		\includegraphics[width=0.95\textwidth]{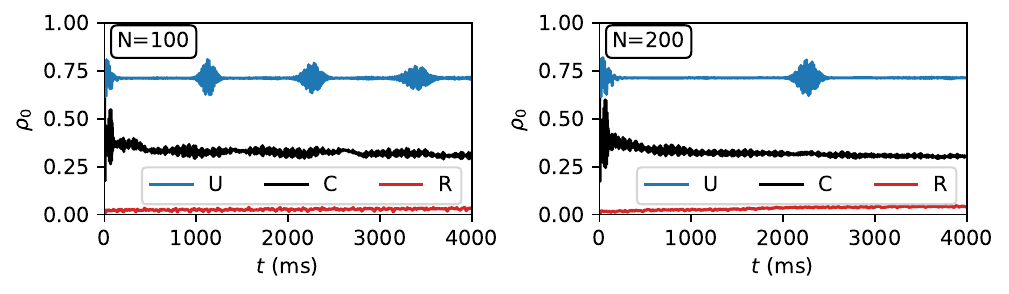}
		\caption{Dynamics of the fractional population $\rho_0$ for the $\textit{R} $,  $\textit{C} $ and  $\textit{U} $ s states for $p=2.0$ Hz for $N=100$ (left)  and $N=200$ particles (right). \label{SI-fig:revivals}}
	\end{figure}
	
	One aspect of the UPO state that is hard to experimentally observe is the eventual return to its initial state due to the periodic nature of its orbits (at least in the semi-classical limit). 
	Figure~\ref{SI-fig:revivals} shows that the UPO initial state, in contrast to the regular and chaotic states, demonstrates a revival also in the dynamics of expectation values,  with a period that scales with the number of particles as seen when comparing the left and right panels. This provides a further distinction even in the non-thermal phase between the regular QMBS and the scarred UPO state. We note that for the experimental number of particles the required time period is significantly beyond any accessible scales.